\documentclass[prl,aps,twocolumn,superscriptaddress,showpacs,floatfix]{revtex4}
\usepackage{graphicx}
\usepackage{subfigure}
\usepackage{amssymb}

\begin{document}

\title{Computing Counterion Densities at Intermediate Coupling}

\author{Christian D. Santangelo}
\email{santancd@physics.upenn.edu}
\affiliation{Department of Physics and Astronomy, University of Pennsylvania, Philadelphia, 19104}
\affiliation{Department of Physics, University of California, Santa Barbara 93106}
\date{\today}

\begin{abstract}
By decomposing the Coulomb interaction into a long distance component appropriate for mean-field theory, and a nonmean-field short distance component, we compute the counterion density near a charged surface for all values of the counterion coupling parameter.  A modified strong-coupling expansion that is manifestly finite at all coupling strengths is used to treat the short distance component.  We find a nonperturbative correction related to the lateral counterion correlations that modifies the density at intermediate coupling.  
\end{abstract}

\pacs{82.70.-y, 82.45.-h,87.15.Kg}

\maketitle

The rise of biological physics has rekindled the long-standing interest in aqueous electrostatics~\cite{electrostaticsreviews}.  Poisson-Boltzmann mean-field theory fails to describe a number of striking phenomena, such as charge inversion~\cite{chargeinversion, wignercrystal} and counterion-mediated attraction~\cite{likechargeattraction, Ha&Liu, bloomfield1, podgornik1, larson1, lau1, lau2}, that occur when strong correlations develop between multivalent counterions.  Although there has been some success understanding counterion correlations using both a phenomenological Wigner crystal theory~\cite{wignercrystal}, and systematic weak-coupling (WC)~\cite{weak-coupling} and strong-coupling (SC)~\cite{strong-coupling,moreira1,moreira} expansions, a complete quantitative theory spanning the entire range of counterion behavior is still lacking.  This letter introduces a method to compute the counterion density at intermediate coupling by decomposing the Coulomb interaction into long and short distance components in the spirit of the Weeks-Chandler-Andersen theory of simple fluids~\cite{WCA}.  This decomposition not only gives good quantitative agreement with simulations, it also provides a natural framework to understand both the success of SC expansion, as well as the role that lateral correlations play in the counterion density.

Here, I use mean-field theory for the long distance interaction, and introduce a modified SC expansion at short distances.  The traditional SC expansion~\cite{strong-coupling} is problematic because it is formally a virial expansion, and one naively expects it to be invalid precisely when the counterions are strongly interacting.  Nonetheless, numerical simulations have demonstrated that it not only correctly predicts the average counterion density in the strong-coupling limit, but also computes the \textit{form} of the corrections~\cite{moreira}.  In contrast, the modified SC expansion introduced here is manifestly finite in the limit of infinite counterion coupling, but recovers the SC corrections at large, finite coupling.

In addition, this decomposition correctly reproduces the counterion distribution around a charged surface at both strong- and weak- coupling, and agrees very well with simulations at intermediate coupling.  A test-charge theory (TCT) which also computes approximate counterion densities at intermediate coupling and explains the exponential form in the strong coupling limit fails to elucidate the physics behind the corrections to that limit in a clear and satisfactory manner~\cite{burak}.  Furthermore, I find a nonperturbative correction to the density related to the lateral counterion correlations that becomes important at intermediate coupling.  Unlike in SC theory, I can unambiguously compute an expression for the free energy and show that these lateral correlations play a role.

Consider the primitive model for a charged surface neutralized by pointlike counterions of the opposite charge in a dielectric medium with dielectric constant, $\epsilon$.  To proceed, introduce a length scale, $\ell$, and define $V_s(\textbf{r}) = l_B Q^2 e^{-r/\ell}/r$ and $V_l(\textbf{r}) = l_B Q^2 (1- e^{-r/\ell})/r$, where $l_B = e^2/(\epsilon k_B T)$ is the Bjerrum length and $Q$ is the counterion valence.  The length $\ell$ is currently arbitrary and will be chosen later to optimize the calculation.  The Hamiltonian for ions of charge $Q e$ centered at positions $\textbf{R}_\alpha$ interacting with a surface of charge density $n_f(\textbf{r}) = \sigma \delta(z)$ is given by
\begin{equation}
\mathcal{H} = \frac{1}{2} \int d^3r d^3r' \rho(\textbf{r}) [V_s(\textbf{r}-\textbf{r}') + V_l(\textbf{r}-\textbf{r}')] \rho(\textbf{r}'),
\end{equation}
where $\rho(\textbf{r}) = n_f(\textbf{r})/Q - \sum_\alpha \delta (\textbf{r}- \textbf{R}_\alpha)$.  It is understood in this expression that ion self-interactions, which arise only from $V_s(\textbf{r})$, are to be neglected.  The long range interaction can be decoupled by introducing a continuous field $\phi$ through a Hubbard-Stratonovich transformation, resulting in the action $S' = S_s + S_l$, where
\begin{eqnarray}
S_s &=& \frac{1}{2} \int d^3r d^3r'~\frac{n_f(\textbf{r})}{Q} V_s(| \textbf{r} - \textbf{r}'| ) \frac{n_f(\textbf{r})}{Q}\nonumber\\
& & - \sum_\alpha \int d^3r~n_f(\textbf{r}) V_s(|\textbf{r}-\textbf{R}_\alpha|)/Q\\
& & + \sum_{\alpha < \beta} V_s( |\textbf{R}_\alpha - \textbf{R}_\beta| ), \nonumber
\end{eqnarray}
$\ell_B = 4 \pi l_B Q^2$, and
\begin{equation}
S_l = \frac{1}{2 \ell_B} \int d^3r [ (\nabla \phi )^2+\ell^2 ( \nabla^2 \phi )^2 ] + i \int d^3r~\frac{\rho(\textbf{r}) \phi(\textbf{r})}{Q}.
\end{equation}
The counterion positions, $\textbf{R}_\alpha$, are restricted to be over the volume of space that can be occupied by the counterions.

In the Grand Canonical ensemble for the counterions, the partition function is
\begin{equation}
Z \propto \sum_N \frac{1}{N!} \left(\frac{\kappa^2}{\ell_B} \right)^N \int \mathcal{D}\phi~\prod_{\alpha=1}^N d^3R_\alpha e^{-S'},
\end{equation}
where the length $\kappa^{-1}$ is defined by $\kappa^2/\ell_B = e^{\beta \mu}/a^3$, $a$ is the counterion radius, and $\mu$ is the chemical potential~\cite{weak-coupling}.  Now define the partial partition function with integrals only over the counterion positions,
\begin{eqnarray}
Z_s &=& \sum_{N=0}^\infty \frac{1}{N!} \int \prod_{\alpha=1}^N \left[ d^3R_\alpha \rho_0(\textbf{R}_\alpha) \right]\nonumber\\
& & \times \exp \left[ - \sum_{\alpha< \beta} V_s ( |\textbf{R}_\alpha - \textbf{R}_\beta|) \right] \label{eq:Zs}
\end{eqnarray}
where $\rho_0 = (\kappa^2/\ell_B) e^{F+i \phi}$, and $F(\textbf{r}) = \int d^3r'~V_s(|\textbf{r}'-\textbf{r}|) n_f(\textbf{r}')/Q$.  This leaves $Z \propto \int \mathcal{D}\phi e^{-S}$, where
\begin{equation}\label{eq:action}
S = \frac{1}{\ell_B} \int d^3r~\left[\frac{1}{2} \left(\nabla \phi \right)^2 + \frac{\ell^2}{2} \left( \nabla^2 \phi \right)^2 \right] - \ln Z_s \left[ \rho_0(\textbf{r}) \right].
\end{equation}
The average counterion density can be formally computed as $\langle \rho(\textbf{r}) \rangle = \delta \ln Z_s/ \delta F(\textbf{r})$.

Since this formulation is exact, the partition function is independent of the choice of $\ell$.  To proceed, make the following approximations: (1) the mean-field approximation for the long-distance interaction (saddle point in $\phi$), and (2) expand the effective potential $\ln Z_s[ \rho_0]$ using a modified SC-like cluster expansion described below.   Making these approximations, the theory will lose its independence on the choice of $\ell$, and there will be a ``best'' $\ell$ whose value gives the closest agreement with the full theory.  In principle, its value should be determined by optimizing the error between a loop expansion on the long-distance interaction and perturbative corrections to the short-distance expansion.  On physical grounds, however, we argue that $\ell \sim l_B Q^2$, or, in other words, $\ell$ is the distance that fixed counterions interact with an energy of $k_B T$.  Counterions at separations larger than this interact weakly, and mean-field theory is likely valid above this length scale.  In addition, the distinction between short and long length scales should not depend on the geometry of the fixed charge distribution, and can therefore only depend on the Bjerrum length and counterion valence.  For concreteness, I will choose $\ell = l_B Q^2$.

The mean-field approximation, given by $\delta S/\delta \phi = 0$, results in the equation
\begin{equation}
\label{eq:meanfield}
\nabla^2 \phi - \ell^2 \nabla^4 \phi + \ell_B \langle \rho(\textbf{r}) \rangle = n_f(\textbf{r}) \ell_B/Q,
\end{equation}
where I have performed the Wick rotation $\phi \rightarrow i \phi$ in the complex plane for convenience.  The SC expansion can be reproduced by expanding $\ln Z_s$ in powers of $\rho_0(z)$.  However, for counterions in the presence of a charged surface with charge density $\sigma$, each term of this expansion diverges as the coupling constant $\Gamma = 2 \pi l_B \sigma Q^3 \rightarrow \infty$ indicating that the counterion interactions are not small; this divergence can be absorbed by shifting $\kappa^2$ and utilizing overall charge neutrality~\cite{strong-coupling}.

Instead, expand in powers of $\delta \rho_0(z) = \rho_0(z) - 2/(\ell_B \lambda) \delta(z)$, where $\lambda= (2 \pi \sigma l_B Q)^{-1}$ is the Gouy-Chapman length for a surface of charge density $\sigma$.  This has the property that $\int dz~\delta \rho_0(z) = 0$ due to overall charge neutrality, and yields
\begin{eqnarray}\label{eq:Zsfull}
Z_s &=& \sum_n \frac{1}{n!} \int \prod_{\alpha=1}^n d^2r_\alpha dz_\alpha \prod_\alpha \delta \tilde{\rho}_0(\textbf{r}_\alpha,z_\alpha)\\
& & \times \exp \left[ - \sum_{\alpha < \beta} V_s(\textbf{r}_\alpha - \textbf{r}_\beta; z_\alpha - z_\beta) \right] Z_p,\nonumber
\end{eqnarray}
where $\textbf{r}_\alpha$ indicates the position of a counterion projected to the surface, $z_\alpha$ its distance from the surface, and $\delta \tilde{\rho}_0(\textbf{r}_\alpha,z_\alpha) = \langle \exp [-\sum_{i} V_s (\textbf{r}_\alpha - \textbf{r}_i;z_\alpha) ] \rangle_p \delta \rho_0(\textbf{r}_\alpha,z_\alpha)$.
Here, $\langle \cdots \rangle_p$ is the average taken with respect to the partition function
\begin{equation}
Z_p = \sum_m \frac{1}{m!} \left( \frac{2}{\ell_B \lambda} \right)^m \int \prod_{i=1}^{m} d^2r_i \exp \left[ - \sum_{i<j} V_s(\textbf{r}_i - \textbf{r}_j;0) \right].
\end{equation}
The expression $\langle \exp [-\sum_{i} V_s (\textbf{r}_\alpha - \textbf{r}_i;z_\alpha) ] \rangle_p$ represents the interaction of a charge at coordinates $(\textbf{r}_\alpha, z_\alpha)$ with a layer of counterions at positions $(\textbf{r}_i,z_i=0)$.  In deriving Eq. (\ref{eq:Zsfull}), I have assumed $\langle \prod_\alpha \exp [-\sum_{i} V_s (\textbf{r}_\alpha - \textbf{r}_i;z_\alpha) ] \rangle_p \approx \prod_\alpha \langle \exp [-\sum_{i} V_s (\textbf{r}_\alpha - \textbf{r}_i;z_\alpha) ] \rangle_p$, which is true as long as the counterions at $z>0$ are far enough apart compared to $\ell$.

Performing a cluster expansion with respect to $\delta \tilde{\rho}_0$ yields $\ln Z_s[\phi] = \int d^3r~\delta \tilde{\rho}_0(\textbf{r}) [ 1 +\frac{1}{2} \int d^3r'~v_2(\textbf{r}-\textbf{r}') \delta \tilde{\rho}_0(\textbf{r}') + \cdots ]$, where $v_2(\textbf{r}) = 1-\exp \left[ - V_s(\textbf{r}) \right]$.  Terms higher than zeroth order vanish in the limit $\Gamma \rightarrow \infty$ as the density becomes delta function-like and $\delta \tilde{\rho}_0 \rightarrow 0$.  As $\Gamma \rightarrow 0$, these corrections also vanish because $\ell^3 \delta \tilde{\rho}_0 \rightarrow 0$, and the interactions become predominantly long-ranged.

It is useful to define $\tilde{\rho}_0 = e^{\zeta(z) - \phi(z)}$ with $\exp[\zeta(\textbf{r},z)] = \langle \exp[F(z) -\sum_i V_s(\textbf{r}-\textbf{r}_i; 0)] \rangle$.  The function $\zeta(z)$ can be interpreted as the short-distance interaction potential of a charge at height $z$ with the charged surface and with a layer of counterions at $z=0$.  Therefore, it encodes the response of the $z=0$ layer to the presence of a charge at some $z>0$, and is reminiscent of theTCT~\cite{burak}.  One difference between $\zeta(z)$ and the TCT, however, is that $\zeta(z)$ also depends, at least in principle, on the short-range structure of the counterions induced by the short-range interaction.  To develop a simple approximation for $\zeta(z)$ which I will use throughout the remainder of the letter, assume that each counterion at $z>0$ interacts with a uniform distribution of charge at $z=0$ containing an induced circular correlation hole of radius $r_0 = \sqrt{\ell_B \lambda/2}=\sqrt{Q/\sigma}$.  This approximates the size of a vacancy in a locally ordered lattice of counterions at the surface, which should be valid when $\Gamma$ is large.  Thus, $\zeta(z) = (\ell/\lambda) [e^{-z/\ell} - e^{\sqrt{r_0^2 + z^2}/\ell}]$.  It is interesting that, for $z \ll \ell$, $\zeta(z) \approx \Gamma (1-e^{-r_0/\ell}) - z/\lambda$ is dominated by the interaction of the counterions with the bare surface and not the $z=0$ layer of counterions whose contribution is of order $\sim z^2/(\lambda \ell) \ll z/\lambda$.

To lowest order, $\langle \rho \rangle = \tilde{\rho}_0$, and the mean-field equation for a charged surface now reads
\begin{equation}\label{eq:mPBeqn}
\partial_z^2 \phi - \ell^2 \partial_z^4 \phi + \kappa^2 \Theta(z) e^{\zeta(z) - \phi(z)} = \frac{\ell_B \sigma}{Q} \delta(z),
\end{equation}
where $\sigma$ is the surface charge density.  For small $z$, $\zeta(z)-F(z)$ is analytic and can be expanded as a power series in $(z/\ell)^2$.  On the other hand, $F(z)$ is nonanalytic at $z=0$ and contributes to the boundary conditions.  This additional contribution can be disentangled by defining $\Phi = \phi - F$.  In terms of $\Phi$, $\zeta(z)$ is replaced with $\zeta(z)-F(z)$ and there is an additional source term on the right of Eq. (\ref{eq:mPBeqn}) of the form $- \ell_B \sigma \ell^2 \partial_z^2 \delta(z)/Q$.
This equation encodes two boundary conditions: $\partial_z \Phi |_{0-}^{0+} - \ell^2 \partial_z^3 \Phi |_{0-}^{0+} = \ell_B \sigma/Q$, and $\partial_z \Phi |_{0-}^{0+} = \ell_B \sigma/Q$.  In terms of the original $\phi$, the boundary conditions are $\partial_z \phi |_{0-}^{0+} = 0$, and $- \ell^2 \partial_z^3 \phi |_{0-}^{0+} = \sigma \ell_B/Q$.  Charge neutrality, $\int dz~\kappa^2 \exp[\zeta(z) - \phi(z)] = \ell_B \sigma/Q$, is ensured for any solution to Eq. (\ref{eq:mPBeqn}).

Analytical approximations to Eq. (\ref{eq:mPBeqn}) can be found in the WC and SC limits.  In the WC limit, I assume the solution will decay with characteristic length $\lambda$.  Thus, the fourth order derivative is negligible.  Since $\zeta(\textbf{r}) \ll 1$, $\phi$ has the PB equation form, given by
\begin{equation}\label{eq:PBsoln}
\phi(z>0) = 2 \ln \left( 1 + \kappa z/\sqrt{2} \right).
\end{equation}
The boundary conditions are satisfied by choosing $\phi(z<0) = 2 (\ell/\lambda) A (e^{z/\ell}-1)$, where $\ell^2 A^3/\lambda^2 - A = 1$.  This requires $\kappa \lambda/2 = A$.  As $\Gamma \rightarrow 0$, $\phi(z<0) \rightarrow 0$, and Eq. (\ref{eq:PBsoln}) becomes exact.

In the SC limit, the fourth order term dominates over the second order term near the surface.  I make the additional assumption that $\phi \ll 1$, and solve
\begin{equation}
\ell^2 \partial_z^4 \phi = \kappa^2 e^{\zeta(z)} \approx \kappa^2 e^{\zeta(0) - z/\lambda},
\end{equation}
which has the solution 
\begin{equation}
\phi(z>0) = \kappa^2 e^{\zeta(0)} \left( e^{-z/\lambda}-1 \right).
\end{equation}
Applying the boundary conditions, $\phi(z<0) = - (2/\ell^2)/(1-\lambda^2/\ell^2) [\exp(z/\ell)-1]$ and $\kappa^2 = (2/\lambda^2)/(1-\lambda^2/\ell^2) \exp[-\zeta(0)]$.  The exponential SC density is recovered for large $\Gamma$.  This solution also becomes exact as $\Gamma \rightarrow \infty$.

For $z \gg \ell$, $\zeta(z) \ll 1$ and Eq. (\ref{eq:mPBeqn}) has an approximate solution given by Eq. (\ref{eq:PBsoln}).  The fourth order derivative can be neglected in this limit because $\kappa$ is exponentially suppressed by $\zeta(0)$ being large. Therefore, the density at large distances is PB-like and is controlled by a renormalized Gouy-Chapman length, $\lambda_{ren} = \sqrt{2}/\kappa = \lambda \exp[\zeta(0)]$.  This is in agreement with the arguments of Burak \textit{et al.}~\cite{burak}, which also exhibits a crossover to a slow decay far from the surface.  Using $\zeta(z)$, I obtain the estimate $\ln (\lambda_{ren}/\lambda) = \Gamma [1-e^{-\sqrt{\ell_B/(2 \ell \Gamma)}}]$.

\begin{figure}[t]
\begin{center}
\resizebox{3in}{!}{\includegraphics{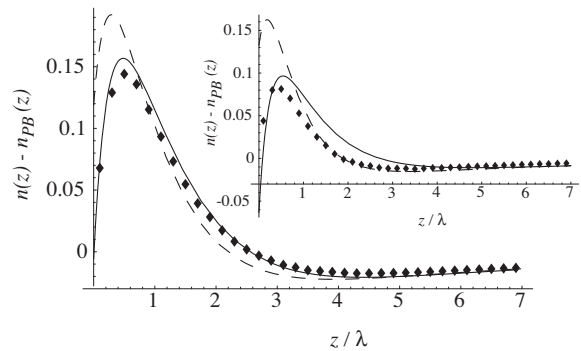}}
\caption{Normalized density difference, $n(z)-n_{PB}(z)$, as a function of $z/\lambda$ for $\Gamma=100$ ($\Gamma=10$ in inset).  Solutions to Eq. (\ref{eq:mPBeqn}) (solid line) are compared to numerical simulations from Ref.~\cite{moreira} (diamonds) and the TCT from Ref.~\cite{burak} (dashed line).}
\label{fig:densities}
\end{center}
\end{figure}

Eq. (\ref{eq:mPBeqn}) has been solved numerically for $\Gamma=100$ and $\Gamma = 10$ (inset).  Fig.~\ref{fig:densities} plots $n(z)-n_{PB}(z)$, where $n(z) = \rho(z) \ell_B \lambda^2/2$ and $n_{PB}(z) = 1/(1+z/\lambda)^2$ is the normalized Poisson-Boltzmann density.  These numerical solutions are compared to actual simulation data from Ref.~\cite{moreira} (courtesy of A.~Moreira) and show quite good agreement.  Furthermore, Eq. (\ref{eq:mPBeqn}) outperforms the TCT (shown as dashed lines using data provided by Y.~Burak from Ref.~\cite{burak}).  The nonperturbative function $\zeta(z)$ is an important component of this numerical agreement; when a virial expansion in $\rho_0$ is used to compute $\ln Z_s$ (equivalent to setting $\zeta(z) = F(z)$ at lowest order), the agreement with the simulation data is only slightly better than the TCT and not nearly as good as Fig.~\ref{fig:densities}.  A more careful evaluation of $\zeta(z)$ is likely to improve these results further.  To be clear, the value for $r_0$ used in Fig.~\ref{fig:densities} is simply an estimate of the real correlation hole size, which may differ up to a factor of order one depending on the model used.  Though there is still very good agreement for other reasonable values of $r_0$, $r_0 = \sqrt{\ell_B \lambda/2}$ seems to give the best agreement with simulations.  In contrast to the $\Gamma=10$ and $100$ results, the density at $\Gamma=1$ always decays faster than the Poisson-Boltzmann density (not shown).  This is precisely where both the short and long-distance expansions in the interpolation scheme attain their maximum error, and it is possible that including higher order terms may improve this.

The SC expansion can be reconstructed in this framework as an asymptotic expansion around the $\Gamma=\infty$ limit by considering the higher order terms in $\delta \tilde{\rho}_0$ for $\ln Z_s$.  Here we sketch this result: substituting the conjectured asymptotically exact result $\tilde{\rho}_0 = 2 e^{-z/\lambda}/(\ell_B \lambda^2)$ into $\langle \rho \rangle$, the first order correction can be computed.  It is straightforward to show that this gives only the \textit{finite} part of the first order SC correction as $\Gamma \rightarrow \infty$, and therefore that this correction vanishes in this limit~\cite{strong-coupling}.  This occurs because the delta function in $\delta \tilde{\rho}_0$ exactly cancels the divergence from $\tilde{\rho}_0$ as $\Gamma \rightarrow \infty$.  Higher order terms will also vanish in this limit, and an asymptotic expansion can be constructed in powers of $1/\Gamma$ which agrees exactly with SC up to first order and which I conjecture also agrees at all orders.  It is interesting that, although no explicit renormalization is necessary in the modified expansion, $\zeta(z)$ arises to shift the fugacity $\kappa^2$ in a similar manner as the SC renormalization.  A more systematic accounting of these corrections is left for future work.

Despite the quantitative agreement, the modified SC expansion suggests a different physical picture of the strong-coupling limit: the corrections to the SC limit arise from the interactions between only those counterions that have made excursions away from the wall, as measured by $\delta \tilde{\rho}_0$.  It is clear that the density of these excited counterions becomes small for large $\Gamma$, and that the SC limit becomes exact.  The function $\zeta(z)$ encodes the interaction of these excitations with their $z=0$ correlation holes, and becomes important at intermediate coupling, once counterions get far enough from the surface.   Interestingly, this correlation hole picture has already been described in the context of Wigner crystal-like structural correlations for multivalent ions~\cite{wignercrystal, nguyen} and in the interpretation of the TCT~\cite{burak}.

This scheme also provides a natural framework to compute the counterion free energy, which is obscured by the traditional SC expansion~\cite{strong-coupling}.  This is given by $F = S(i \phi) - \mu N/(k_B T)$, where $N$ is the number of counterions, $\phi$ is the mean-field long range potential, and the chemical potential is related to $\kappa^2$ by $\mu = k_B T \ln (\kappa^2 a^3/\ell_B)$.  Since $\kappa^2$ depends exponentially on $\zeta(0)$, nonperturbative correlations also play a role in the free energy, and subsequently in the interaction between two surfaces at separations where the SC expansion cannot be applied.  This will be explored in a future publication.

To summarize, I have computed the counterion density around a charged surface using a scheme to decompose the Coulomb interaction into short and long distance components.  Each is treated with different approximations.  For large $\Gamma$, we recover the SC results and for small $\Gamma$ we recover the WC Poisson-Boltzmann density.  At intermediate coupling, the model agrees very well with the simulation data, and depends crucially on a nonperturbative correlation correction whose form we have estimated.  These correlations also play a role in determining the renormalized Gouy-Chapman length when the densities recovers its Poisson-Boltzman form far from the surface.

\begin{acknowledgements}
  I would like to thank A.~Moreira for providing the simulation data and Y.~Burak for providing solutions for the TCT.  I also acknowledge many helpful discussions with P.~Pincus, M.~Henle, D.~Needleman, A.W.C.~Lau, A.~Liu, and R.~Kamien. This work was supported by the NSF under Award
  No.~DMR02-03755, Award No.~DMR01-29804, and by the Materials Research Laboratory at UCSB under Award No.~DMR00-80034.
\end{acknowledgements}

\end{document}